# Smith-Purcell Radiation from Compound Blazed Gratings

B. P. Clarke[1], J. K. So[1], K. F. MacDonald[1] and N. I. Zheludev[1, 2]

[1] *Optoelectronics Research Centre & Centre for Photonic Metamaterials, University of Southampton, Southampton, SO17 1BJ, UK*

[2] *Centre for Disruptive Photonic Technologies, School of Physical and Applied Sciences & The Photonics Institute, Nanyang Technological University, Singapore 637371*

Free electrons travelling in vacuum carry evanescent electric field, which can only be coupled to free space light via interaction with a surrounding or nearby medium: as Cherenkov radiation when travelling faster than the local velocity of light or as "Smith-Purcell" (SP) radiation when passing over the surface of a grating. The SP emission characteristics of simple gratings such as a regular array of wires or slits are a well-understood phenomenon. Here we show that for a compound grating made up of several closely-spaced slits repeated for each period, the characteristic angular dispersion of Smith-Purcell radiation can be selectively attenuated or enhanced. We analyze, in particular, the change in intensity as slit elements are added to the structure for two slit sizes and how this affects emission enhancement.

Impact and proximity interactions between nanostructured materials and medium-energy free electrons have been harnessed to remarkable effect recently in various forms of electron-induced radiation emission (EIRE) imaging [1–4]. These techniques rely upon the spectroscopic, emission angle-, and polarization-resolved analysis of visible to near-infrared light emitted when electrons, with energies of a few tens of keV (velocities of a few tenths of the speed of light, typical of scanning electron microscopes) strike or pass within nanoscale range of a target to probe and spatially map the photonic and plasmonic properties of nanomaterials[5–7]. Short-pulse (usually femtosecond laser-driven) electron sources add the dimension of temporal resolution[8,9].

The same underlying mechanisms of interaction among free electrons, photons and nanostructured materials have also been engaged for the realization of microscopic electron-beam-driven, optical frequency light sources. A variety of periodic structures from simple planar gratings to cylindrical metal/dielectric 'undulators', photonic crystals, metamaterial and nanoantenna arrays, and 2D holographic gratings have been engineered to couple electron energy to free-space and guided light modes of prescribed, if not tunable, wavelength, direction, divergence and topological charge[1,2,10,11].

It is well-established in diffractive optics that blazed and complex, compound gratings can provide for manipulation by design of the spectral and spatial distribution of reflected, transmitted and guided light modes by adding new degrees of freedom for control of the phase distribution of electromagnetic field. They find application in such areas as metrology[12], beam shaping[13,14], spectral filtering[15], spectroscopy and sensing[16]. Gradient metasurfaces provide for similarly refined control of reflection/transmission phase distributions in the domain of sub-wavelength periodicity, enabling for example polarization-independent reflection[17], achromatic flat optical elements[18], and nonlinear optical repsonses[19,20]. Similar principles may be applied to tailor the electron-induced light emission from grating structures (Fig. 1). Here we computationally analyze optical frequency light emission from compound gratings excited by the passage of medium-energy free electrons travelling parallel to the grating plane. It is found that by varying the effective 'fill factor' of a multi-slit Smith-Purcell grating, the emission efficiency of the grating at both the fundamental and higher-order wavelengths can be modified, and even enhanced over a simple rectangular grating.

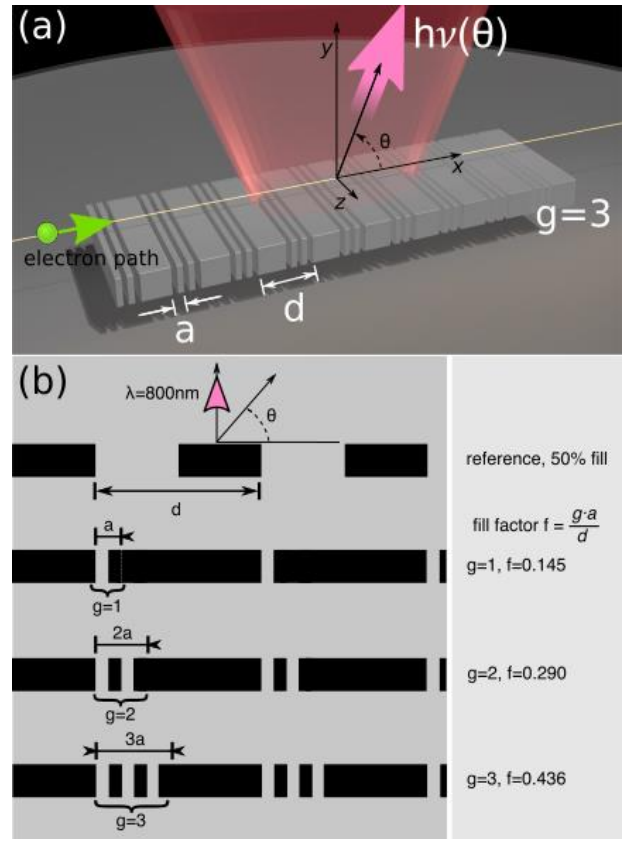

**Figure 1. Compound grating Smith Purcell light sources:** (a) Artistic impression of electron-beam-induced optical frequency radiation emission from a grating comprised of $g$ [= 3 as shown] slots of width $a$ within each fundamental period of length $d$. (b) Dimensional schematic of numerically simulated compound grating geometries [illustrated for values of $g$ up to 3], and the simple rectangular reference grating.

Free electrons carry time-varying electromagnetic fields, spanning a wide spectral range that extends to high optical frequencies[21–23].This field decays exponentially on the sub-optical-wavelength scale (except in the case of highly relativistic electrons[24,25], and remains evanescent unless the speed of light in the surrounding medium is less than the electron velocity, in which case Cerenkov radiation will be generated[26], or unless additional momentum is provided via scattering at an optical inhomogeneity as electrons pass close to (within a few tens of nanometers of) a material structure. In the particular case of a grating, the resulting light emission is known as diffraction[24] or Smith-Purcell (SP) radiation[25,27]. The SP mechanism is well-understood and has been has been extensively studied for its potential in generating particularly terahertz radiation (not readily obtained by other means), including in the context of 'Smith-Purcell free electron lasers' driven by pulsed, high-energy (relativistic) electron beams[28]. It has been applied recently to the realization of optical fiber probes that sample electron evanescent fields in a manner reminiscent of scanning near-field optical microscopy probes for evanescent light fields – coupling electron energy to propagating UV/visible guided modes in the fiber[29], and to the demonstration of electron evanescent field amplification via a plasmonic "poor man's superlens"[3].

A simple grating of period $d$ provides additional parallel momentum in integer multiples ($n$) of its reciprocal lattice vector, $2\pi/d$, to incoming electromagnetic waves, leading to familiar optical diffraction phenomena for light propagating in free space. The same applies in the SP emission process to free electron evanescent fields, enabling a part of the electromagnetic energy to be decoupled into propagating waves when the parallel momentum is smaller than that of light in free space. The angular dispersion of SP emission is given by $\lambda = (d/n)(\beta^{-1} - \cos\theta)$, where $\lambda$ is the emission wavelength, $\beta = v/c$ is the velocity of electron relative to the speed of light, and $\theta$ is the angle between the direction of light emission and the electron propagation direction[27].

In the THz domain, it has been proposed that improvements in monochromaticity and directivity may be achieved by modifying the fill factor[30,31] of simple, singly periodic Smith-Purcell gratings, by graduating their depth[32], or by making them aperiodic[33]. In the present case, we analyze SP emission at optical frequencies from a family of compound grating structures using the particle-in-cell finite-difference time-domain (PIC-FDTD) computational method in CST Particle Studio, whereby both the electron trajectory and electro-dynamic response of gratings can be simulated.

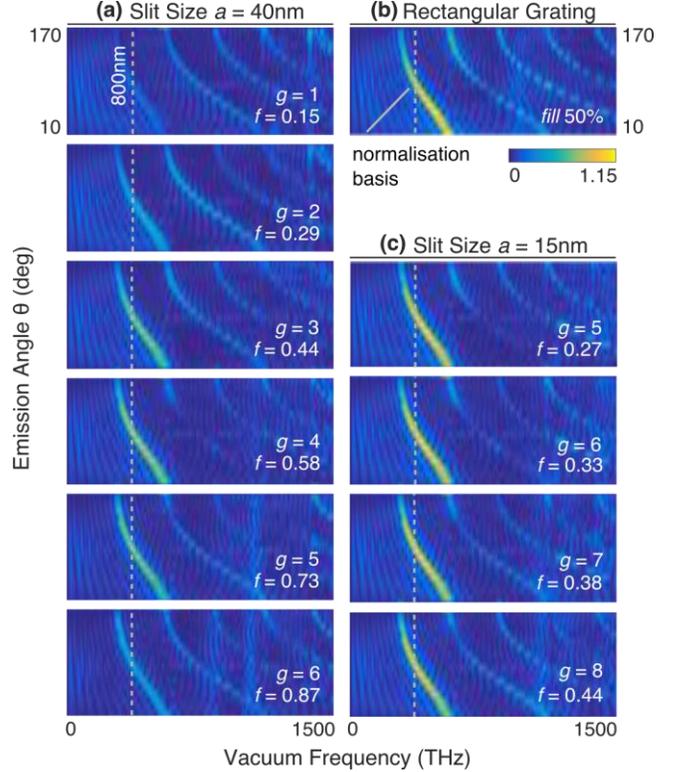

**Figure 2. Smith-Purcell emission from compound gratings:** Angular dispersion of first and second order [$n$ = 1, 2] SP emission wavelength for gratings [simple or compound] of period $d$ = 275 nm at an electron energy $E$ = 32.5 keV for: (a) compound gratings of slit element size $a$ = 40nm, up to $g$ = 6 slits. (b) Rectangular grating of 50% fill used as normalization basis. (c) compound gratings of slit element size $a$ = 15nm, up to $g$ = 8 slits, showing enhancement of 7% over the normalization basis for $\theta$ = 90°, $\lambda$ = 800nm.

We consider 2D geometries in the $xz$ plane as illustrated in Fig. 1b: gratings are modelled as 50nm thick layers of perfect electrical conductor in vacuum, perforated with patterns of vertical slits. Each comprises 12 periods of length $d$, within which there are $g$ slits of width $a/2$ and (sub-)periodic spacing $a$. We define a grating fill factor $f = ga/d$. The transient electrodynamic simulation space is bounded by perfectly matched layers in the $z$ directions at $z = \pm 500$nm and in the $x$ direction at the grating ends. The field is then propagated 100μm radially to calculate the far field result at each angle $\theta$. Electrons propagate in the $+x$ direction with a beam width of 4 nm (corresponding to a minimum of 2 meshing cells in the vacuum region). Our reference structure is a simple rectangular grating for which $f$ = 0.5 (i.e. $g$ = 1; $a$

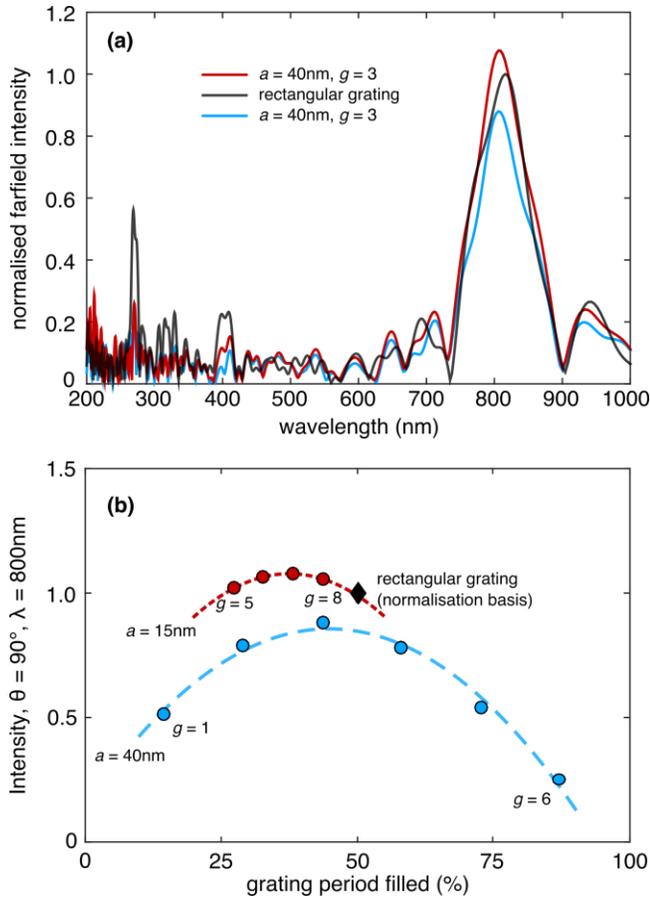

**Figure 3. Compound Smith-Purcell grating efficiency:** (a) Spectrum of surface-normal [$\theta = 90°$] far-field intensity for the most efficient compound SP gratings at each slit dimension used: slit widths $a = 40$ nm, $g = 3$, (blue line). and $a = 15$nm, $g = 7$, (red line). Shown normalized to the peak 800nm intensity for a rectangular reference grating (black line). (b) Dependences of 800 nm, normal direction far-field intensity on effective fill factor, i.e. the number of sub-periodic slots $g$, for slot widths $a = 15$ nm (red dashed line) and 40 nm (blue dashed line). The single filled diamond marker represents the reference grating. In all cases, the fundamental period d = 275 nm and electron energy e = 32.5 keV.

$= d$). We initially identify a combination of period $d = 275$ nm, electron energy $E = 32.5$ keV, and impact parameter (distance between grating surface and electron beam axis) $h = 7$ nm that optimizes light emission at $\lambda = 800$ nm in the surface-normal ($\theta = 90°$) direction for the reference grating. These values are maintained for all subsequent iterations as the sub-periodic structure is varied.

For this grating geometry, the 'effective fill factor' of the compound-grating model was increased as shown in Fig. 1, by increasing the number of slits, $g$, from 1 to 6, the maximum possible for a period of $d = 275$nm and a silt distance of $a = 40$nm.

With the same fundamental period $d$, compound gratings maintain the same relationship between emission wavelength $\lambda$ and emission angle $\theta$ as the reference grating for emission orders $n = 1$ and $n = 2$, as shown in Fig. 2. The relative magnitudes of emission (c.f. diffraction efficiencies) though are functions of the sub-periodic parameters $a$ and $g$. For a fixed slit width of 20 nm ($a = 40$ nm):

Surface-normal emission at 800 nm is maximized when $g = 3$, reaching an intensity ~85% of that of the simple reference grating. This maximum is attained when the fill factor is closest to 0.5, i.e. closest to that of the reference structure, though for smaller slot widths the optimum fill factor is found to decrease (Fig. 3).

Stronger second-order emission ($n = 2$; $\lambda = 400$ nm) at $\theta = 90°$ is observed from the compound gratings than from the simple rectangular reference grating, when $g = 5$. Second-order emission intensity is maximized at this grating number and angle, an enhancement of 34% over the rectangular grating. Increased levels of second order emission over the rectangular grating model are also observed at shallower angles relative to the electron trajectory for other grating numbers. For example, for $g = 2$, $\theta = 70°$, an enhancement of 24% is seen. This can be seen as analogous to the behavior of compound optical diffraction gratings at varying angles. As the incident angle is swept for a given wavelength, the reflection efficiency generally increases, but with several null points dependent on the geometry. At many particular angles, the reflection efficiency can be greater than that of a standard rectangular grating[34].

Following the increase in surface-normal emission as the effective fill factor approached $f = 0.5$ in the $a = 40$nm gratings, a series of gratings with reduced slit size was tested to see whether emission could be enhanced above that of a standard Smith-Purcell effect rectangular grating of 50% fill. For each element size, different numbers of sub-element slits were used to find the most efficient grating possible for 800nm emission normal to the surface. While computational limits provide a lower bound on the size of the grating studied, a modest enhancement of 7% over the standard rectangular grating was observed for a slit distance of $a = 15$, with 7 gratings, as seen in Fig. 3. For this slit size, with $g = 5$ to 8, the spectrum is also shown in Fig. 2. These gratings all show decrease in intensity of the higher-order (n = 2) mode.

In conclusion, we have demonstrated a compound Smith-Purcell grating with enhanced emission in the normal direction when compared to a standard rectangular grating. Furthermore, the principles of compound and multi-element diffraction gratings can be used as a tool in understanding Smith-Purcell radiation from repeating structures, and in shaping these emissions to suit a variety of purposes. A better understanding of these structures can yield applications in free-electron lasing[35], narrow-band radiation sources[30], and efficient terahertz emitters[32].

This work was supported by the Engineering and Physical Sciences Research Council, UK [Projects EP/M009122/1 and EP/G060363/1], and the Singapore Ministry of Education [grant MOE2011-T3-1-005].